\documentclass{article}
\usepackage[reqno,intlimits]{amsmath}
\usepackage{amssymb}
\usepackage[round]{natbib}
\usepackage[dvips]{graphicx}
\usepackage[english]{babel}
\usepackage{comment}
\usepackage{hyperref} 
\newcommand{\quotes}[1]{``#1''}

 \title{CPT violations in Astrophysics and Cosmology}

 \author{G. Auriemma\\Universit\`{a} degli Studi della Basilicata, Potenza, Italy\\
and\\INFN Sezione di Roma "La Sapienza", Roma, Italy}
\date{June, 2007}

\begin{document}
\maketitle
\begin{abstract}
   In this paper it is given a brief review of the current limits
on the magnitude of CPT and Lorentz Invariance violations, currently
predicted in connection with quantum gravity and string/M-theory,
that can be derived from astrophysical and cosmological data. Even
if not completely unambiguous, these observational tests of
fundamental physics are complementary to the ones obtained by
accelerator experiments and by ground or space based direct
experiments, because potentially can access very high energies and
large distances.
\end{abstract}
\section{Introduction}\label{sect:intro}
\cite{Einstein1905AnP322891E} introduced the postulate of the
constancy of the velocity of light in empty space, justifying it on
the bases of the negative result of the Michelson-Morley experiment
\citep{michelson1887}. Since then the covariance of physical laws
under the group of Lorentz transformations, usually nicknamed the
``Lorentz invariance''(LI), has been proven to hold locally within a
very high accuracy in tests done on the Earth or nearby space. A
modern version of the Michelson-Morley experiment has been performed
comparing the frequency of solid state resonant cavities, in two
orthogonal mode while in rotation respect to the Cosmic Microwave
Background Radiation (CMBR) reference frame. Nevertheless the
accuracy of the measurements is of the order of $10^{-7}$ for
resonators in motion with the Earth
\citep{SaathoffPhysRevLett.91.190403} and of $10^{-10}$ for rapidly
spinning ones \citep{Stanwix:2006jb}. As we will discuss briefly in
the remaining part of this introduction, this level of accuracy
cannot exclude that Special Relativity Theory could be only an
approximation of the physical reality, that could be violated by
tiny perturbations introduced by conceivable mechanism.\par Around
the middle of the last century, the development of a fully
Relativistic Quantum Field Theory (RFQT) has pointed out  that there
is a strong logical connection between the Lorentz invariance and
the matter and antimatter duality. This is not unexpected since the
very existence of antimatter is a consequence of the
relativistically covariant Dirac equation \citep{Dirac1933nl}.
Following an early indication by \cite{SchwingerQFT-I},
\cite{Pauli1955book} introduced the idea that for every process
occurring in nature there is an allowed one, occurring with the same
probability, in which each particle is replaced with its
corresponding antiparticle, having reversed spin and following a
trajectory that is the reflection under space and time inversion of
the original ones. This is now called the ``Schwinger-Pauli-Luders
CPT theorem'' because \cite{Luders1957} showed that assuming
\begin{enumerate}
\item Lorentz Invariance;
\item Locality;
\item Unitarity;
\item Correct connection between spin and statistics
\end{enumerate}
it is possible to prove that the successive application of charge
conjugation C, parity reflection P and time reversal T in any order
leave the Hamiltonian invariant. As a complement
\citet{Greenberg2002prl} has also proved the \quotes{inverse-CPT
theorem}, showing that if under any circumstances CPT invariance is
violated, then also Lorentz invariance will be.\par In certain sense
the CPT invariance is more fundamental than Lorentz invariance,
because, as we will show later, it is possible to conceive
modifications of the Hamiltonian of the fundamental interactions
which violates Special Relativity, but are invariant under CPT
transformations, acting equally on particles and antiparticles.\par
The CPT theorem points out what could be the physical origin of
Lorentz violations. In particular the second hypothesis that leads
to it is \quotes{locality}, which demands that space-like separated
events should not influence each other \citep{Wess:1988dm}. There
are at least three good theoretical reasons to suspect that locality
could not hold for arbitrary small distances:
\begin{enumerate}
    \item  The unavoidable singularities of the General Theory of Relativity (GRT)
\citep{Hawking1982} that makes the structure of space-time very
complex at distances of the order of the Planck length
$\ell_P=\sqrt{\hbar G/ c^3}\simeq1.66\times 10^{-35}\hbox{ m}$. The
fabric of space-time has been vividly described as a \quotes{foamy}
structure (see \emph{e.g.} \citealt{Wheeler1998book}), turbulently
perturbed by black holes continuously popping out of the vacuum and
evaporating in times of the order of $\Delta t= c\ell_P$.\bigskip
    \item String/M-theories  are intrinsically non local theories \citep{Amati:1988tn}
    where the ordinary concept of point-like components of matter is
    substituted by two dimensional object with a finite, even if
    very small, dimensions . The space-time structure of the string
    theory is discontinuous on a scale $\Delta t\Delta x\ge c
  \ell_s^2$ , being $\ell_s$ the size of the string
\citep{Yoneya:1989ai}. If string theory has incorporates gravity,
one of the characteristic length of the theory should be
$\ell_s^{grav}=\ell_P$, but might exist other characteristic
lengths, corresponding to different type of interactions, with
$\ell_s>\ell_P$ \citep{Lykken:1996fj}.
\bigskip
\item In string theory, gravity is just one of the many possible excitations of a string
(or other extended object) living over some background metric space.
The existence of such background metric space, over which the theory
is defined, is needed for the formulation and for the interpretation
of the theory. The \quotes{Loop Quantum Gravity} theory
\citep{lrr-1998-1} is an attempt to eliminate this background
space-time. In this theory the space-time has a kind of
\quotes{polymeric} structure with minimal space cell with volumes
$\Delta V\sim \ell_p^3$ \citep{Rovelli:1994ge}.
\end{enumerate}
In any case the indeterminacy principle states that to resolve a
length scale $\ell$ we need energies of the order of $\Lambda=\hbar
c/\ell=(\ell_P/\ell)M_P$ where $M_P=\sqrt{\hbar c/G}=1.22\times
10^{19}\;\mathrm{GeV}$ is the Planck mass. Na\^{\i}ve expectations
of the orders of magnitude for $CPT$ violations, motivated by
dimensional considerations, will be
\begin{equation}\label{eq:intro1}
    \Delta\mathcal{H}\approx \frac{E^2}{\Lambda}
\end{equation}
\par At low energies the Hamiltonian is of the order of
magnitude of the mass of the particle, therefore CPT symmetry can be
tested in the laboratory measuring the difference of mass between
particles and antiparticles. The best constraint has been obtained
from the limits on the mass difference between the neutral strange
mesons $K^0-\overline{K}^0$ recently obtained from the KLOE
experiment at the Daphne $\varphi$-factory
(LNF-INFN)\citep{Ambrosino:2006ek}
\begin{equation}\label{eq:intro2}
    \frac{|m_{K^0}-m_{\overline{K}^0}|}{m_{K^0}}<1.26 \times
10^{-18}\quad \mathrm{(95\%\; C.L.)}
\end{equation}
that from Eq.\ \eqref{eq:intro1} is predicted to be $\approx
0.5\times 10^{-19}$.\par Astrophysical and cosmological test of the
CPT/LI violations, that I will discuss in the rest of this paper are
complementary to local tests, because test modifications of physical
law over large spatial scale $D\sim10^{26}\hbox{ m}$ and long times
$t_0\sim13.7\hbox{ Gy}$.
\section{Parameterizations of the CPT/LI
violations}\label{sec:param} Many different theoretical frameworks
for CPT/LI have been investigated in detail (see e.g.
\citealt{Mattingly:2005re} for a recent review). The closer to
physical intuition is the ``Modified Dispersion Relations'' (MDR)
framework, that has the advantage of supplying a relatively model
independent parametrization of the CPT/LI, at the expense of rigor
and completeness. Nevertheless this approach is, in my opinion for
an experimentally oriented paper.\par Assuming that the CPT/LI
violating Hamiltonian of a free particle or field can be written
$\mathcal{H}=\mathcal{H}_{free}+\Delta\mathcal{H}$ where
$\Delta\mathcal{H}$ is a small perturbation of the standard
Hamiltonian $\mathcal{H}_{free}=\sqrt{p^2+m^2}$, whose order of
magnitude will be given by Eq.\ \eqref{eq:intro1}, we can put
\begin{equation}\label{eq:param1}
    E^2-p^2-m^2=F(E,\vec p)
\end{equation}
where the R.H.S. term of this equation can be expanded in Taylor
series as
\begin{equation}\label{eq:param2}
   F(E,\vec p)=F^{(1)}_\mu p^\mu+ F^{(2)}_{\nu\rho} p^\nu
p^\rho+F^{(3)}_{\sigma \kappa \lambda}\,p^\sigma p^\kappa
p^\lambda+\cdots
\end{equation}
where $p^\mu\equiv\{E,\vec p\}$ is the four momentum.  It is worth
noticing that we can derive many qualitative feature of CPT/LI
violations from this relatively simple \emph{Ansatz}.
\begin{itemize}
\item  From a phenomenological point of view there is no \emph{a priori} reason to
expect that the coefficients in Eq.\ \eqref{eq:param2} are
universal, even if the fundamental Lorentz violation is universal
(for a discussion on this assumption see \emph{e.g.}
\citealt{Alfaro:2004aa}). At least we expect from  Eq.\
\eqref{eq:intro1} a dependance from the energy, that at low momentum
is a dependence from the mass. In general we must assume an implicit
dependence from all the conserved quantum numbers of the particle,
namely intrinsic spin, charges, flavor, etc.
\bigskip
 \item Only the odd terms of Eq.\ \eqref{eq:param2} are CPT
violating (\quotes{CPT odd}) while even terms are CPT conserving
(\quotes{CPT even}), therefore we have the relation:
\begin{equation}\label{eq:param3}
    \overline{F}^{(n)}_{\mu_1\mu_2\cdots \mu_n}=(-1)^n
F^{(n)}_{\mu_1\mu_2\cdots \mu_n}
\end{equation}
where obviously $\overline{F}^{(n)}$ is the coefficient for the
corresponding free antiparticle.\bigskip
\item Moreover the odd terms violate also P and T conjugation.
This make a distinction between the right-handed and the left-handed
component of a particle (or field) with spin, because under P and T
the four-momentum of the field changes direction while the spin
conserve its direction. In practice we will have
\begin{equation}\label{eq:param4}
F^{(n)}_{\mu_1\mu_2\cdots k_\mu}=\zeta^n |F^{(n)}_{\mu_1\mu_2\cdots
\mu_n}|
\end{equation}
where $\zeta=\pm1$ will be the polarization index of the particle,
with $\zeta=+1$ if the spin is $\vec s\uparrow\uparrow\vec p$ and
$\zeta=-1$ if $\vec s\uparrow\downarrow\vec p$. This fact induces a
spin precession in the propagation of the particle in vacuum
(birefrangence of the vacuum).\bigskip
\item From dimensional argument we expect
\begin{equation}\label{eq:param5}
    \left|F^{(1)}_k p^k\right|\approx \left|F^{(2)}_{ji} p^j
p^i \right|\approx\left| F^{(3)}_{qrs}p^q p^r p^s\right|
\sim\mathcal{O}\left(\frac{|p|^3}{\Lambda}\right)
\end{equation}
while for $n>3$ will be
\begin{equation}\label{eq:param6}
 \left|F^{(n)}_{\mu_1\mu_2\cdots \mu_n}p^{\mu_1}p^{\mu_2}\cdots p^{\mu_n}
 \right|\sim\mathcal{O}\left(\frac{|p|^n}{\Lambda^{n-2}}\right)
\end{equation} As a
consequence at leading order it will be necessary to consider the
firsts 3 terms of the dispersion relation's expansion, that are
expected to have about the same order of magnitude.\bigskip
\end{itemize}
\section{Tests on a preferential direction in space-time}\label{sect:direct}
The coefficient of the first term on R.H.S. of Eq.\
\eqref{eq:param2} is a four-vector with the dimensions of a mass,
that assign a preferential direction in space-time, sometimes called
in the literature the \quotes{Chern-Simons term}. We can write the
MDR of a photon, including only the directional term in the form
\begin{equation}\label{eq:direct1}
    \omega=\sqrt{k^2+2\zeta_\gamma (\xi_0 \omega-\boldsymbol\xi\cdot \vec k)}
\end{equation}
where we have put $F^{(1)}_{\mu}\equiv
    \{\xi_0/2,-\boldsymbol\xi\}$. Solving this equation we have the explicit
    form of the MDR
$$\omega=\zeta_\gamma\xi_0\pm\sqrt{k^2+\zeta_\gamma^2\xi_0^2-2\zeta_\gamma\boldsymbol\xi\cdot \vec k}
    \simeq \pm k+\zeta_\gamma(\xi_0\mp
    |\boldsymbol\xi|\cos\theta)$$
where $\theta$ is obviously the angle between $\vec k$ and
$\boldsymbol\xi$. The ambiguity of the sign is only apparent because
the lower sign is meaningful only in the case that $k<0$, but in
this case $\cos\theta<0$. Therefore the true physical solution is
\begin{equation}\label{eq:direct2}
\omega\simeq k+\zeta_\gamma(\xi_0-
    |\boldsymbol\xi|\cos\theta)
\end{equation}
As we said before the polarization index $\zeta_\gamma$ is $+1$ for
a right-handed circularly polarized wave and $-1$ for the opposite
polarization, therefore this vacuum birefrangence effect could be
detected in the propagation of polarized radiation. A linearly
polarized wave is represented by the superposition
    of two circularly polarized waves
    \begin{equation}\label{eq:direct3}
        \Psi=\Psi_0\left\{e^{-i\alpha-\omega_+ t}\boldsymbol\epsilon_+
        +e^{i\alpha-\omega_- t}\boldsymbol\epsilon_-\right\}
    \end{equation}
where $\alpha$ is the initial polarization angle. It is evident that
the polarization angle as a function of time will be
$\alpha(t)=\alpha_0+(\omega_+-\omega_-)t$ from which we have that
the polarization plane of radiation emitted by a source at
cosmological redshift $z$ will rotate at Earth by an angle
\begin{equation}\label{eq:direct4}
\Delta\alpha=2\int_0^{z}(\xi_0(z)-|\boldsymbol\xi(z)|\cos\theta)\frac{\,dz}{(1+z)H(z)}
\end{equation}
It is interesting that this rotation would be independent upon the
wavelength of the radiation, being distinguishable from the
interstellar Faraday rotation that is $\propto \lambda^{2}$.\par
\cite{Noland1997} claimed that data on polarized radiation emitted
by distant radio galaxies show a marginal statistical evidence
($3\sigma$) for a systematic rotation depending on the angle
$\theta$ between the propagation wave vector $\vec k$ of the
radiation and a direction roughly localized  in a region
$19h\le\alpha\le23h, -20^\circ\le\delta\le+20^\circ$, that could be
explained by $|\boldsymbol\xi|\simeq 10^{-41}\hbox{ GeV}$. But this
claim was not confirmed by a reanalysis of the same data with
different statistical techniques \citep{Loredo1997}. A search to a
sample of 160 radiogalaxies with $0.3<z<2.12$ \cite{Carrol11997}
found
\begin{equation}\label{eq:direct5}
    \xi_0=(0.8\pm1)\times10^{-41}h_0\hbox{ GeV}\quad |\boldsymbol \xi|=(1.5\pm1.9)\times10^{-41}h_0\hbox{ GeV}
\end{equation}
Independently \cite{Wardle1997} from  observation of the
polarization in the optical V band of 3C265 ($z=0.82$) found a mean
deviation of $-1.4^\circ\pm1.1^\circ$, that yields a limit from Eq.\
\eqref{eq:direct4} at present time, assuming a moderate evolution $
|\boldsymbol\xi|\propto (1+z)$ the upper limit
\begin{equation}\label{eq:direct6}
    |\boldsymbol\xi|\le 2\times10^{-41}\hbox{ GeV}\quad\hbox{(95\%
    C.L.)}
\end{equation}
This limit indicates that a directional term is strongly suppressed,
even respect to the dimensional estimate $(\hbar\omega)^2/M_P\sim
4\times10^{-37}\hbox{ GeV}$.\par
    \cite{Feng:2006dp} speculate that a preferred space-time direction in the Universe could be
   originate by a scalar field $\phi$ that might constitute the so called ``quintessential dark
    energy'' (for a recent review see e.g. \citealt{Copeland:2006wr}). In this
    case in the CMBR rest frame the four-vector $\xi_\mu$ is time-like, being $\xi_0=\dot\phi$.
    We can estimate the order of magnitude of the vector from
    observations of the expansion rate of the Universe using the
    equation \citep{PeeblesPhysRevD1988}
    \begin{equation}\label{eq:direct7}
        \dot\phi^2=\frac{16\pi}{M_P^2}(1+w_\phi)\rho_\phi
    \end{equation}
where $\rho_\phi$ is the dark energy density and
$p_\phi=w_\phi\rho_\phi$ its equation of state. A recent estimate
\citep{Riess:2004nr} sets an upper limit $w\le-0.76$ at 95\% C.L.
from which, assuming $\Omega_V=0.732\pm0.018$
\citep{Spergel:2006hy}, we estimate from Eq.\ \eqref{eq:direct7}
$\xi'_0<10^{-41}\,h_0\hbox{ GeV}$. In the solar system frame, moving
with $V\simeq
    370\;\mathrm{km/s}$ respect to CMBR, the components of the four
    vector would be $\xi_0=\gamma\,\xi'_0$ and
              $ |\boldsymbol\xi|=\gamma\beta \xi'_0\simeq1.2\times
               10^{-3}\xi_0$.

\section{Test of CPT-odd violations from the polarization of the CMBR}\label{sect:prop}
 In the MDR framework, expressed by the
expansion of Eq.\ \eqref{eq:param2} in a space-time isotropic
Universe, neglecting directional effects that appears from
experiments very suppressed, the dispersion relation for photons at
leading order is
\begin{equation}\label{eq:prop1}
   \omega=\sqrt{k^2+\zeta_\gamma\frac{\eta_\gamma}{M_P}k^3}\simeq
   k\left(1+\zeta_\gamma \frac{\eta_\gamma}{M_P}\frac{k}{2}\right)
\end{equation}
where $\eta_\gamma$ is an adimensional parametrization  of the
magnitude of CPT-odd LI violations. From this dispersion relation,
we obtain the phase velocity of light
\begin{equation}\label{eq:prop2}
    c_{\gamma}(\omega,\zeta_\gamma)=\frac{\omega}{k}\simeq1+\zeta_\gamma\frac{\eta_\gamma}{M_P}\frac{\omega}{2}
\end{equation}
where we have preserved the measured value $c_\gamma=1$ for
$\omega\ll M_P$.\par The two circular polarization of the photons
will propagate  with different phase velocity (cosmological
birefrangence)
\begin{equation}\label{eq:prop3}
    \Delta v(\omega)=c_{\gamma}(\omega,+1)-c_{\gamma}(\omega,-1)=\frac{\eta_\gamma}{M_P}\omega
\end{equation}
As in the case of directional term, illustrated in the previous
\S\ref{sect:direct}, the plane of polarization of a linearly
polarized wave from a source at redshift $z$ is rotated of an angle,
that in this case, substituting
\begin{equation}\label{eq:prop4}
    \omega(z)=\frac{2\pi c_\gamma}{\lambda }(1+z)
\end{equation}
where $\lambda$ will be the detected wavelength of the photon, will
be
\begin{equation}\label{eq:prop5}
    \Delta\alpha(z)\simeq\frac{2\pi}{M_P}\lambda^{-1}\int_{0}^{z}\frac{\eta_\gamma(z)}{H(z)}dz
\end{equation}
We observe that the adimensional CPT-odd coefficient is expected to
be by dimensional argument $\eta_\gamma\propto \hbar\omega/\Lambda$,
making the effective dependence of the rotation angle $\propto
\lambda^{-2}$, distinguishable from the interstellar Faraday
rotation that is as we said before $\Delta\alpha_{F}\propto
\lambda^{2}$.\par The detailed maps of CMBR temperature and
polarization, obtained from WMAP \citep{Page:2006hz}, offer an
intriguing possibility to set limits to the cosmological
birifrangence at redshifts $0\le z \le 1100$ (as proposed earlier by
\citealt{PhysRevLett.83.1506}).
\begin{itemize}
\item From fits of WMAP and BOOMERANG data \cite{Feng:2006dp,Xia:2007qs}
obtain the limit $$\Delta\alpha=-6.2^\circ\pm3.8^\circ$$
\item From wavelet fits of WMAP 3-year data \cite{Cabella:2007br} obtain a rotation of
polarization of the CMBR
$$\Delta\alpha=-2.5^\circ\pm3^\circ$$
\end{itemize}
The rotation of polarization expected from Eq.\ \eqref{eq:prop5},
averaged over the spectrum of the CMBR, assuming an evolution
$\eta_\gamma(z)=\eta_\gamma(0)(1+z)$, is
\begin{equation}\label{eq:prop6}
    \Delta\alpha_{CPT}\simeq
11.7^\circ\,h_0^{-1}\eta_\gamma
\end{equation}
The result of \cite{Cabella:2007br} is close to being a significant
negative experiment, because gives the upper limit
\begin{equation}\label{eq:prop7}
    \eta_\gamma<0.2\,h_0\quad\hbox{(95\% C.L.)}
\end{equation}
that imposes the energy scale of CPT-odd violations for photons to
be $\Lambda \ge5\,h_0^{-1}\, M_P$.\par
\section{Test of CPT-odd violations from GRB's}\label{sect:burst} Unpolarized radiation can be represented by the
superposition of two equal amplitude waves, with opposite circular
polarization. The group velocity of the photon in vacuum will at
leading order will be
\begin{equation}\label{eq:burst1}
    v_\gamma(\omega,\zeta_\gamma)=\frac{\partial\omega}{\partial k}\simeq1+\zeta_\gamma\frac{\eta_\gamma}{M_P}\omega
\end{equation}
slightly different from the phase velocity given by Eq.\
\eqref{eq:prop2}. This introduce a time spread $\Delta t\propto
v_{\gamma}(\omega,+1)-v_{\gamma-}(\omega,-1)$ that is, for a source
at redshift $z$ given by
\begin{equation}\label{eq:burst2}
    \Delta t(z)\simeq\frac{4\pi}{M_P}\lambda^{-1}\int_0^z\frac{\eta_\gamma(z)}{H(z)}dz
\end{equation}
where we will assume as in the previous section
$\eta_\gamma(z)=\eta_\gamma(0)(1+z)$.\par\cite{Amelino-Camelia:1997gz}
suggested that GRB could be used to constraint the vacuum dispersion
of radiation, due to their short intrinsic duration and high energy
emission. However it appears that the present limits that can be
obtained by this method cannot really access the Planck scale. In
fact from Eq.\ \eqref{eq:burst2} we derive a time spread for a burst
in the hard X-ray band
\begin{equation}\label{eq:burst3}
    \Delta t(z=1)\simeq 22.7\left(\frac{\hbar\omega}{200\hbox{ keV}}\right)\eta_\gamma\,h_0^{-1}\;\mu\mathrm{s}
\end{equation}
The shortest time scale ever detected has been observed in the
exceptional GRB920229 \citep{Schaefer1999}, observed by BATSE to
have a rise time $\tau=220\pm30\;\mu\hbox{s}$. From the negligible
time dispersion of the rise of the burst among the low energy
channel (25-50 keV) and the most populated channel (100-300 keV),
that we estimate $<130\;\mu\hbox{s}$, and assuming a redshift
$z\approx 1$ \citep{Amelino-Camelia:1998qp}, we can set a limit
$\eta_\gamma< 5-6\,h_0$ which implies $\Lambda>0.2\,h_0^{-1}\;M_P$.
Similar limits are obtained with an accurate statistical analysis of
a sample of 35 GRB's with known redshift \citep{Ellis:2005wr}.\par
The observation of linear polarization of the prompt emission from
GRB at cosmological distances could set stringent limits to the
birifrangent propagation, that is implied by Eq.\ \eqref{eq:prop2}.
In fact linear polarized $\gamma$-ray photon with different energies
will be rotated by an amount given by Eq.\ \eqref{eq:prop5}.
Therefore the observation of linear polarizations in the prompt
emission of GRB's could give a very strong limit to the vacuum
birifrangence \citep{Mitrofanov2003Natur.426Q.139M}.\par
\citealt{Fan:2007zb} from the observation of the afterglows of
GRB020813 and GRB 021004 in the UV band could set the limit
$|\eta_\gamma|\le 10^{-7}$, if one can rule out an intrinsic origin
for the rotation of the polarization vector at various energies.

\section{Tests on CPT violations from the Crab Nebula}\label{sect:emint} The dispersion
relation of charged particles (electron or protons), in a space-time
isotropic Universe, can be put in the form
\citep{MyersPhysRevLett.90.211601}
\begin{equation}\label{eq:emint1}
    E(p,\zeta_p)=\sqrt{m^2+(1+\epsilon_p) p^2+(\eta'_p+\zeta_p\eta_p)\frac{p^3}{M_P}}
\end{equation}
where the coefficients $\epsilon_p$, $\eta_p$ and $\eta'_p$  are
adimensional, expected to be $\propto M_P/\Lambda$. In this formula
we have introduced a distinction between the C-even part of the
coefficient of cubic modification to the dispersion relation
$\eta'_p$ and its C-odd part $\eta_p$. It is evident from this
expression that the maximum attainable phase velocity of particle,
for $E\ll M_P$, will be
\begin{equation}\label{eq:emint2}
    c_p\simeq\sqrt{1+\epsilon_p}\neq 1
\end{equation}
In the moderate ultrarelativistic regime $m\ll E\ll M_P$ we assume
the C-even parameter $\eta_p'=0$ and we have
\begin{equation}\label{eq:emint3}
    E\simeq p c_p\left(1+\frac{m^2}{2 p^2 c_p^2}+\zeta_p\frac{\eta_p}{2 M_P c_p^2} p\right)
\end{equation}
Deriving this equation and substituting $E\approx p\,c_p$ we find
the group velocity of the De Broglie wave associated to the particle
\begin{equation}\label{eq:emint4}
    v(E,\zeta_p)=\frac{\partial E}{\partial p}\simeq
    1-\frac{m^2}{2E^2}+\zeta_p\frac{\eta}{M_P c_p^2}E
\end{equation}
From this we calculate the energy and helicity dependent Lorentz
factor at leading order
\begin{equation}\label{eq:emint5}
    \gamma(E,\zeta_p)=\frac{1}{\sqrt{1-\frac{v^2}{c_p^2}}}
    \simeq\left(\frac{m^2}{E^2}-2\zeta_p\frac{\eta_p}{M_P c_p^2}
E\right)^{-1/2}
\end{equation}
The peculiarity of this formula is that for $\zeta_p=+1$
(right-handed particles) the Lorentz factor shows an apparent
divergence that is likely canceled by higher order terms, while for
$\zeta_p=-1$ (left-handed particles) it has a maximum value
$\gamma_{max}\simeq1.7\times10^{7}/\eta_p^{1/3}$ for $E\simeq
14.7/\eta_p^{1/3}\hbox{ TeV}$.\par Several observable modifications
of familiar e.m. radiation processes follow from this fact, as we
will show in the rest of this section, can be understood easily from
kinematical considerations. We begin with the Compton scattering
\begin{equation}\label{eq:emint6}
    e^\pm+\gamma\to e^\pm+\gamma
\end{equation}
Following a well known method \citep{BLUMENTHALRevModPhys.42.237} we
consider the scattering as occurring in the Thomson limit
$\tilde\omega'\simeq \tilde\omega$ in the rest frame of the
electron, where $\tilde\omega$ and $\tilde \omega'$ are the incoming
and outgoing energy of the photon. In order to calculate these
energies in the electron rest frame, we consider that from Eq.\
\eqref{eq:prop1} follows that, also in presence of CPT/LI violations
we have
\begin{equation}\label{eq:emint7}
    \omega^2-k^2 c_\gamma^2(\omega,\zeta_\gamma)=0
\end{equation}
in any inertial reference. Condition that is realized by the
pseudo-Lorentz transformations:
\begin{equation}\label{eq:emint8}
    \omega'=\omega
    \frac{c_\gamma-v\cos\theta}{\sqrt{c_\gamma^2-v^2}}\quad ;\quad
    k'=\frac{\omega'}{c'_\gamma }\quad ;\quad \tan\theta'=\frac{\sqrt{c_\gamma^2-v^2}\sin\theta}{c_\gamma\cos\theta-v}
\end{equation}
Using these transformations we have the energy of the incoming and
outgoing photons in the electron rest frame:
\begin{equation}\label{eq:emint9}
    \tilde\omega=\omega
    \frac{c_\gamma(\omega,\zeta_\gamma)-v\cos\theta}{\sqrt{c_\gamma^2(\omega,\zeta_\gamma)-v^2}}\quad;\quad
    \omega'=\tilde\omega'\frac{c_\gamma(\tilde\omega',\zeta_\gamma)-v\cos\theta'}{\sqrt{c_\gamma^2(\tilde\omega',\zeta_\gamma)-v^2}}
\end{equation}
The maximum energy in the laboratory of the scattered photon will be
approximately
\begin{equation}\label{eq:emint10}
    \omega^{max}_{IC}\simeq\omega\frac{c_\gamma(\omega')+v}{c_\gamma(\omega)-v}
\end{equation}
that tends to the well known standard expression
     $\omega^{max}_{IC}=\omega\gamma^2(1+\beta)^2$ in the Lorentz invariant
     limit.\par
 We observe that the corrections to the nominator of Eq.\ \eqref{eq:emint10} are negligible,
 while they determine the order of magnitude of the denominator. In
 practice we will put $c_e=1$ (see later in this section for a justification of this assumption) and use
 the approximation
\begin{equation}\label{eq:emint11}
    \omega^{max}_{IC}\simeq
4\omega\left(\frac{m^2}{E^2}+\zeta_\gamma\frac{\eta_\gamma}{M_p}\omega-2\zeta_e\frac{\eta_e}{M_p}E\right)^{-1}
\end{equation}
Viewing the magnetic field as a collection of virtual photons with
average energy $\omega_B=e B/m$ that are scattered by the fast
electrons (see \emph{e.g.} \citealt{Lieu1993ApJ}), we can apply the
theory of inverse Compton scattering in presence of Lorentz
violations outlined above. Therefore the maximum energy of the
synchrotron photons will be given by Eq.\ \eqref{eq:emint11} in the
form
\begin{equation}\label{eq:emint12}
    \omega^{\max}_{Sync}\simeq 4\omega_B\left(\frac{m^2}{E^2}
    +\zeta_\gamma\frac{\eta_\gamma}{M_p}\omega_B-2\zeta_e\frac{\eta_e}{M_p}E\right)^{-1}
\end{equation}
In the general case $\omega_B\ll E$ we can neglect the effect of
CPT-odd violations of photon propagation, that is constrained from
Eq.\ \eqref{eq:prop7} to be $\eta_\gamma\le0.14$ (assuming
$h_0=0.7$).  Neglecting a quantity $\le2\times10^{-33}$ the term in
parenthesis at the R.H.S. of Eq.\ \eqref{eq:emint12} is exactly
equal to the Lorentz factor of the electron given by Eq.\
\eqref{eq:emint5}.\par As we have noted above this formula has a
peculiar behavior for left-handed particle, but in usual conditions
the electrons are a mixture of left-handed and right-handed
components. However for massless particle the helicity is a good
quantum number, therefore when $E\gg m$ electrons(positrons) are
almost all left-handed(right-handed) like neutrinos(antineutrinos).
Following \cite{Jacobson:2002ye} we maximize the equation
\begin{equation}\label{eq:emint13}
    \omega^{max}_{Sync}=\frac{e B}{E}\left(\frac{m^2}{E^2}+2\frac{\eta_{e^-}}{M_P}E\right)^{-\frac{3}{2}}
\end{equation}
taking $E$ as an independent variable. The maximum is attained for
an energy of the electron $E=10/\eta^{1/3}\;\mathrm{TeV}$ with a
Lorentz factor $\gamma(E)=1.58\times10^7/\eta^{1/3}$, that gives the
constraint
\begin{equation}\label{eq:emint14}
    \eta_{e^-}\le\frac{M_P}{m}\left(\frac{0.35 e B}{m\,\omega^{max}_{Sync}}\right)^{\frac{3}{2}}
\end{equation}
The Crab Nebula is an excellent astrophysical laboratory for testing
this type of CPT-odd violations. Assuming that synchrotron emission
contribute to the $\gamma$-ray unpulsed spectrum of the Crab Nebula
up to a maximum energy of $\approx100\hbox{ MeV}$
\cite{Jacobson:2002ye} derive a constraint to the CPT-odd violation
parameter
\begin{equation}\label{eq:emint15}
   \eta_{e^-}\le
7\times10^{-8}\left(\frac{\omega^{max}_{Sync}}{100\;\mathrm{MeV}}\right)^{-3/2}\left(\frac{B}{0.6
\; \mathrm{mG}}\right)^{3/2}
\end{equation}
It is worth noticing that this limit is rather conservative, because
the EGRET spectrum of the unpulsed component of the Crab Nebula
\citep{EGRET1993ApJ} has a break at $\sim 1\hbox{ GeV}$, suggesting
a limit $\sim 30$ times smaller. Nevertheless we must point out that
the limit obtained is consistent only if would be possible to
exclude that the synchrotron radiation is emitted by $e^+ e^-$
pairs, because in this case the limits applies only to the electron
component.\footnote{After the presentation of this paper an extended
analysis of the effect of LV on the Crab Nebula emission has been
discussed by \citealt{Maccione:2007yc}, with interesting
results.}.\par No better limit can be obtained from the cutoff in
the UHE part of the spectrum of the nebula, observed by the HESS
experiment at $14.3\pm2\hbox{ TeV}$ \citep{Masterson2005ICRC}. The
emission of the nebula above $\sim 1\hbox{ GeV}$ is extremely well
consistent \citep{deJager1992ApJ,Atoyan1996MNRAS} with the expected
flux produced by Compton scattering of synchrotron, dust IR, and/or
CMBR photons by the same electrons that produce the synchrotron
component. In fact if we apply Eq.\ \eqref{eq:emint11} derived
above, we would expect in any case $\omega^{max}_{IC}>>100\hbox{
TeV}$. Very likely the cause of the $\sim 10\;\mathrm{TeV}$ cut-off
is to be searched in the absorption process $\gamma+\gamma\to
e^++e^-$ that is very effective when $\omega_{IC}\omega_{IR}\ge m^2$
\citep{Telnov:1989sd}.\par In the derivation of the Eq.\
\eqref{eq:emint11} we have neglected $\epsilon_e$. It is worth
noticing that from the UHE emission from The Crab Nebula we can set
also a stringent limit to the CPT-even LI violations. We consider
the process
\begin{equation}\label{eq:emint16}
       \gamma \to e^+ +e^-
\end{equation}
namely pair creation in vacuum. This process is forbidden by LI
because the conservation of four-momentum imposes
\begin{equation}\label{eq:emint17}
    \omega^2-k^2=(E_+^2-p_+^2)+2(E_+E_--p_+p_-\cos\theta)+(E_-^2-p_-^2)
\end{equation}
where the L.H.S. is null and the R.H.S. is $>2m^2$. If we use the
MDR of Eq.\ \eqref{eq:emint1} including only the CPT-even term, Eq.\
\eqref{eq:emint17} is written
\begin{equation}\label{eq:emint17b}
    \omega^2-k^2=2m^2+\epsilon_e(p_+^2+p_-^2)+2(E_+E_--p_+p_-\cos\theta)
\end{equation}
that can be satisfied if $\omega=k$ when $\epsilon_e<0$. The
threshold for the occurrence of the reaction \eqref{eq:emint16} in
presence of CPT-even/LI violations is
\begin{equation}\label{eq:emint17c}
      \omega\ge\frac{2 m_e}{\sqrt{-\epsilon_e}}
\end{equation}
The fact that photons with $E_\gamma>20\;\mathrm{TeV}$ have been
observed from the Crab Nebula set the constraint
\begin{equation}\label{eq:emint18}
   |c_e^2-1|<2.5\times10^{-15}
\end{equation}
comparable with the limit inferred from observations of VHE
$\gamma$-rays from the extragalactic source MK501
\citep{Stecker:2001vb}.
\section{Global tests of special relativity}\label{sect:CR}
As we have discussed before, the leading order CPT-even term in the
expansion of Eq.\ \eqref{eq:param2} induces violation of Special
Relativity that could be observed experimentally, because the
limiting velocity for massive particle is changed by Eq.\
\eqref{eq:emint2} \citep{Coleman1999}.\par An important QED process
is the \v Cerenkov emission in vacuum by protons
    \begin{equation}\label{eq:CR1}
       p \to\gamma +p
    \end{equation}
This process is also forbidden in vacuum by LI because the
four-momentum conservation imposes
\begin{equation}\label{eq:CR2}
    E^2-p^2=E'^2-p'^2+2(\omega E'-k p'\cos\theta)
\end{equation}
but if LI holds the L.H.S of this equation is $E^2-p^2=m_p^2$ while
the r.h.s. is always $>m_p^2$.  On the contrary if LI is violated
from Eq.\ \eqref{eq:emint1} we have
\begin{equation}\label{eq:CR3}
    \epsilon_p(p^2-p'^2)=2(\omega E'-k p'\cos\theta)
\end{equation}
that has solutions for $\epsilon_p>0$. Therefore the  reaction\
\eqref{eq:CR1} occurs over the threshold
\begin{equation}\label{eq:CR4}
    p>\frac{2 m_p}{\sqrt{3\epsilon}}
\end{equation}
 From the simple facts that HiRes \citep{Abbasi:2007sv} and AGASA
 \citep{TakedaPhysRevLett.81.1163} has observed primary cosmic rays with energies up to
 $E_p\sim 10^{19}\;\mathrm{eV}$ we can derive the constraint
\begin{equation}\label{eq:CR2b}
c_p^2-1<1.3\times10^{-20}
\end{equation}
\cite{GreisenPhysRevLett.16.748} and \cite{Zatsepin1966JETPL}
published independently a calculation of the interaction of protons
with the CMBR that should prevent cosmic rays with energy $\ge
10^{20}\;\mathrm{eV}$ to cross distance $d\le 50\;\mathrm{Mpc}$ (the
so called GZK cut-off). The experimental situation is not clear at
the moment, because the data from the HiRes experiment
\citep{Abbasi:2007sv} show an evidence at 5$\sigma$ for the presence
of a sharp cut-off at $6\times 10^{19}\hbox{ eV}$ while the AGASA
data \citep{TakedaPhysRevLett.81.1163} show a flattening (ankle) of
the spectrum above $10^{19}\hbox{ eV}$, reconstructing six events
with energy $\ge 10^{20}\hbox{ eV}$.\par Hopefully the experimental
situation will be resolved in one way or the other by the Pierre
Auger Observatory, expected to be fully operational in few months
from now, because \cite{Glashow:1999zk} showed that UHE cosmic rays
can play an important role in connection with tests of Special
Relativity.\par
 The reaction producing the GZK cut-off is the photoproduction
 reaction
 \begin{equation}\label{eq:CR5}
    p+\gamma\to \Delta^{+}
 \end{equation}
where $\Delta^+$ is the lowest pion-nucleon resonance with a mass
$m_\Delta=1232\hbox{ MeV}$. The four momentum conservation
gives
\begin{equation}\label{eq:CR6}
    E_p^2-p_p^2+2(\omega E_p -k\,p_p\cos\theta)=E^2_\Delta-p^2_\Delta
\end{equation}
Using the MDR expansion in the ultrarelativistic regime $E_p\gg m_p$
we have
\begin{equation}\label{eq:CR7}
    m_p^2+\epsilon_p
    p_p^2+2\omega\,p_p(\sqrt{1+\epsilon_p}-\cos\theta)=m_\Delta^2+\epsilon_\Delta
    p_\Delta^2
\end{equation}
In the LI case $\epsilon_p=\epsilon_\Delta=0$ the threshold for this
reaction is
\begin{equation}\label{eq:CR8}
    p_p\ge \frac{m_\Delta^2-m_p^2}{4\omega}\simeq 5\times
    10^{19}\;\mathrm{eV/c}
\end{equation}
perfectly consistent with the observed cut-off. However we can
approximate $p_\Delta\simeq p_p$ because $k<<p_p$ therefore Eq.\
\eqref{eq:CR7} can be written
\begin{equation}\label{eq:CR9}
    (\epsilon_p-\epsilon_\Delta)p_p^2+2\omega p_p\le
    m_\Delta^2-m_p^2
\end{equation}
that can be satisfied only if
\begin{equation}\label{eq:CR10}
    \epsilon_\Delta-\epsilon_p<\frac{\omega^2}{m_\Delta^2-m_p^2}\simeq
\times10^{-25}
\end{equation}
 This shows that a tiny LI violation can eliminate the GZK cut-off.
 It is clear at this point that a secure confirmation of the
 observation of the GZK cut-off would set a very strong limit to the
 validity of Special Relativity. In fact from the dimensional
 estimate we expect $\epsilon_\Delta-\epsilon_p\sim
 m_\Delta-m_p/M_P\sim2.4\times10^{-20}$, more then 4 order of magnitudes
 larger then the limit that could be inferred from the confirmation
 of the GZK cut-off.\par
\section{Conclusions}
Astrophysical tests show that Einstein's Special Relativity Theory
is in quite good health, far beyond the Planck scale. The final
assessment of the evidence for GZK cut-off in the primary UHE cosmic
ray spectrum would secure an upper limit to CPT-even violations
$\lesssim 10^{-25}$.\par The Crab Nebula promises to be an excellent
particle physics laboratory for the search of CPT-odd effects, that
could allow the exploitation of electron beam energies up to
$\approx 2500\hbox{ TeV}$, but unfortunately the interpretation of
data is far from being lacking in ambiguity. For example if it could
be demonstrated that synchrotron emission is produced by negative
electrons only, a limit to CPT-odd violations of the order of
$\lesssim 10^{-8}\ell_P$ could be assessed.\par The existence of a
preferred direction in space-time, possibly connected with a
quintessential dark-energy, is constrained by the optical
polarimetry of far distant galaxies to be very small $<5\times
10^{-5}\ell_P$, but the scale of anisotropy estimated from the dark
energy density is $<2.5\times 10^{-5}\ell_P$, close but not
conclusive. It is intriguing that in the future X and $\Gamma$-ray
polarimetry of bright objects at cosmological distances, like AGN
and GRB, could improve the present limits, by order of magnitudes,
if emission models are also improved.\par The polarization of the
CMBR is another interesting source of data on possible CPT-odd
violations in the photon sector, but at present the results of WMAP
allow to constraint the scale in the range $\sim 0.1\ell_P$. The
predicted sensitivity of the Planck satellite, to be launched about
one year from now, can improve significantly the above limit in the
next decade.
\bibliographystyle{hapj}
\bibliography{QFT}
\end{document}